\documentclass[sigchi]{acmart}
\usepackage{geometry} 
\usepackage{soul} 
\usepackage{enumitem} 

\usepackage{subcaption} 
\usepackage{float} 

\newenvironment{researchqs}{%
  \begin{list}{\textbf{RQ\arabic{enumi}:}}{%
      \usecounter{enumi}
      \setlength{\leftmargin}{5pt}
      \setlength{\labelwidth}{0pt}
      \setlength{\labelsep}{0.5em}
      \setlength{\itemindent}{0pt}
      \setlength{\itemsep}{0.5\baselineskip}
  }
}{\end{list}}

\usepackage{dirtytalk} 
\AtBeginDocument{%
  }

\copyrightyear{2026}
\acmYear{2026}
\setcopyright{cc}
\setcctype{by}
\acmConference[CHI EA '26]{Extended Abstracts of the 2026 CHI Conference on Human Factors in Computing Systems}{April 13--17, 2026}{Barcelona, Spain}
\acmBooktitle{Extended Abstracts of the 2026 CHI Conference on Human Factors in Computing Systems (CHI EA '26), April 13--17, 2026, Barcelona, Spain}
\acmDOI{10.1145/3772363.3799014}
\acmISBN{979-8-4007-2281-3/2026/04}

\begin{document}

\title[Understanding Clinician Experiences with Game-Based Interventions for Autistic Children]{Understanding Clinician Experiences with Game-Based Interventions for Autistic Children to Inform a Future Game Platform Focused on Improving Motor Skills}

\author{Hunter Beach}
\email{hmb433@nau.edu}
\orcid{0009-0004-0063-2646}
\affiliation{%
  \institution{Northern Arizona University}
  \city{Flagstaff}
  \state{Arizona}
  \country{USA}
}

\author{Devin Jay Debita San Nicolas}
\email{dds324@nau.edu}
\orcid{0009-0006-5575-2754} 
\affiliation{%
  \institution{Northern Arizona University}
  \city{Flagstaff}
  \state{Arizona}
  \country{USA}
}

\author{Carly Miller}
\email{crm764@nau.edu}
\orcid{0009-0001-2857-4773}
\affiliation{%
  \institution{Northern Arizona University}
  \city{Flagstaff}
  \state{Arizona}
  \country{USA}
}

\author{Cathy Ly}
\email{cl2493@nau.edu}
\orcid{0009-0005-3779-8110}
\affiliation{%
  \institution{Northern Arizona University}
  \city{Flagstaff}
  \state{Arizona}
  \country{USA}
}

\author{Jared Duval}
\email{Jared.Duval@nau.edu}
\orcid{0000-0002-5125-0566} 
\affiliation{%
  \institution{Northern Arizona University}
  \streetaddress{1900 S Knoles Dr.}
  \city{Flagstaff}
  \state{Arizona}
  \country{USA}
  \postcode{86011}
}

\renewcommand{\shortauthors}{Beach et al.}

\begin{abstract}
Motor challenges are prevalent among autistic children, and games are able to simultaneously produce clinically meaningful results and provide a motivating context, but many current solutions are too rigid. We conducted a two-phase qualitative study comprised of semi-structured interviews and participatory design workshops with 7 pediatric physical and 5 occupational therapists (PTs/OTs) to investigate their perspectives and experiences with game and play-based interventions. We identified 8 prominent themes describing key characteristics of current successful interventions, opportunities, and barriers to adoption in clinical practice. We present a speculative design informed by thematic analysis that addresses current challenges of rigidity in Serious Games for Health (SG4H). Our modular platform (\textit{AutMotion Studio}) hosts a suite of interventions as customizable minigames, allowing community members to contribute to and employ Wizard of Oz paradigms for flexible appropriation strategies.
\end{abstract}

\ccsdesc[300]{Human-centered computing~Accessibility}
\ccsdesc[500]{Human-centered computing~User models}
\ccsdesc[300]{Human-centered computing~Interaction design}

\keywords{physical therapy, occupational therapy, serious games for health, motor skills, autism}

\maketitle

\section{Introduction}

Motor skills strongly influence quality of life \cite{heller_circus_2018}, yet up to 87\% of autistic individuals experience motor challenges\cite{Zampella2021,norvilitis_sensory_2012, marquez_segura_circus_2019, bundy_sensory_2002}. Evidence shows that early interventions can improve motor capabilities, as well as other developmental factors, including social skills \cite{Zampella2021, jared_duval_reimagining_2023}. Responding to critiques of rigid medical interventions\cite{spielAgencyAutisticChildren2019, kayali_design_2018, baltaxe-admonyDREEMMovingEmpathy2024, duval_chasing_2021, morozovEverythingClickHere2013, charlton_nothing_2000, ymousAmJustTerrified2020, wongResistanceHopeEssays2018} and inspired by work on ambiguity\cite{turmo_vidal_ambiguity_2024}, Wizard of Oz paradigms \cite{dow_wizard_2005, duval2022playful, jared_duval_reimagining_2023}, and customization \cite{duval_designing_2022, morris_broadening_2010, ahsen_designing_2022, 10.1145/3342285, gualano_invisible_2023, zhang_its_2022, 10.1145/3308561.3354637}, we conducted semi-structured interviews with clinicians to explore adoption factors and barriers with 4 occupational therapists and 5 physical therapists. Our needfinding served as a precursor to our participatory design workshops \cite{bodker_participatory_2018, muller_participatory_2002, McIntyre2008, ellis_participatory_1998, MALINVERNI2017535} with 2 occupational therapists and 3 physical therapists. Our research agenda is focused on play because it is an integral part of human development, health, education, and eudaimonia across the lifespan \cite{tekinbas_rules_2003, sicartPlayMatters2014, sicart_playing_2015, ryan_self-determination_2000, rusch_making_2017, walz_gameful_2015, deci_hedonia_2008, GranicIsabela2014TBoP}, making it a key component in pediatric therapy practice \cite{Rautenbach03042025, LopezNieto2022, Kuhaneck2013, peng_need_2012, heller_circus_2018, gobel_serious_2010, hilton_exergaming_2015, rafiei_effects_2021, Jimenez2022}. In recognition that further participatory design work with autistic youth and their families is crucial future work, the research questions related to clinician perspectives that guided this work are as follows:

\begin{researchqs}
        \item \textbf{Design Principles:} \textit{What key requirements, opportunities, challenges, and design considerations should guide the development of effective video game–based interventions for pediatric motor skill therapy among autistic youth?} 
        \item \textbf{Adoption Factors:} \textit{What concrete design particulars do therapists envision to adopt new game interventions for pediatric motor skill therapy for autistic youth?}
\end{researchqs}

Key findings of our needfinding phase include eight themes necessary in interventions: (1) modular task breakdown, (2) flexibility, (3) therapeutic specificity, (4) accessibility, (5) sensory sensitivity (6) skill carry-over beyond clinics, (7) child agency, and (8) social connection. Our participatory design workshops surfaced several insights and concrete design particulars \cite{waernActivityUltimateParticular2017}, including (1) maps that provide high-level goal-setting, (2) Wizard of Oz paradigms via secret video game controller input to accommodate a plurality of bodies, movement patterns, and therapeutic contexts, (3) multimodal user experiences that can be played on varying screen sizes, and (4) designs that provide content for both clinical and home contexts. This work makes the following two contributions:
\begin{enumerate}
    \item \textit{Theory}: An elucidation of themes and design particulars that inform the design of clinically-relevant games to build motor skills among autistic youth
    \item \textit{Artifact}: A speculative design of \textit{AutMotion Studio}, a modular platform that hosts a suite of intrinsically rewarding and clinically-relevant games for motor skill development among autistic youth that values contextual appropriation, embraces intentional ambiguity, and celebrates the autistic lived experience 
\end{enumerate}

This paper is structured as follows: Section \ref{sec:Method} provides an overview of how this research was conducted, Section \ref{sec:Results} details our thematic analysis of semi-structured interviews, Section \ref{sec:Design} presents our speculative design informed by our thematic analysis, Section \ref{sec:Discussion} hosts our insights on the findings, limitations, and future work, and Section \ref{sec:Conclusion} is a concise synthesis of our contributions. 

\section{Method}\label{sec:Method}
\noindent\paragraph{Positionality and Language Use} 

There is no consensus among disability advocacy, academic, and medical/service communities regarding the use of identity-first (e.g., an autistic person) and person-first (e.g., a person with autism) language, or the terms used to describe functional abilities (e.g., “level of function” versus “support needs”) in interventions. Following the preferences of autistic and disabled members of our research lab, this paper uses identity-first language when referring to autistic people, person-first language for broader disability populations, and the term “support needs” when describing functional requirements. However, when directly quoting interviewees, the paper preserves the language they used.

Given that the positionality fundamentally impacts the social dynamics of participatory design, we take space here to describe our positionality. Duval provided oversight and conducted a training on how to facilitate participatory design sessions. The interviews and design sessions were facilitated by Hunter, Devin, and Carly. Three team members identify as male, and two as female. We embrace an intersectional feminist approach to design. Three of the researchers are in the Computer Science program, and one is in the Psychology program.

\noindent\paragraph{Participants} To address \textit{RQ1}, we interviewed nine clinicians (P01-P09) in the fields of pediatric physical and occupational therapy. To address \textit{RQ2}, we then met with two prior interviewees and three additional clinicians (P01, P08, P10-P12) to create a speculative design via participatory design. We acknowledge there is a participant overlap between phases. However, this overlap was limited and did not substantially bias the results as the participatory design sessions involved additional clinicians and focused on generating new speculative design ideas rather than validating prior findings. Recruitment was carried out by contacting clinicians and professors within our professional networks, and then snowball sampling \cite{goodman_snowball_1961}. We collected limited demographic information to minimize participant burden. Participants were licensed pediatric physical (n = 7) and occupational therapists (n = 5), with professional experience ranging from early-career to senior clinicians. Each had experience working across home, clinic, and/or school settings, and all reported regularly working with autistic children.

\noindent\paragraph{Study Design} 
This study is part of a larger Research through Design\cite{zimmerman_research_2007, zimmerman_recentering_2022, zimmerman_analysis_2010} project exploring a modular sociotechnical game platform for improving motor skills among autistic youth. All protocols were approved by our institutional review boards (NAU IRB 1957455-5) and (NAU IRB 2346182-3).

For \textit{RQ1}, this study utilized qualitative data collected through semi-structured interviews of pediatric physical and occupational therapists. After completing informed consent, therapists were interviewed over a Zoom meeting, with questions provided in the supplemental materials. Interviews explored the current methods of game and play-based interventions used by therapists, which features are essential for an effective intervention, and which barriers often present themselves when using a game or play-based intervention. All interviews were audio-recorded and transcribed for analysis. The interviews were transcribed and assessed using inductive thematic analysis \cite{Braun01012006}, following a method similar to that used in Lobo et al. \cite{Lobo2021}. Members of the research team individually read each transcript and used inductive open coding \cite{guestAppliedThematicAnalysis2012, Braun01012006, braun_thematic_2019} to find important and recurring ideas to develop into central themes. Researchers then reached a consensus on the most significant themes, based on the consistency of similar coded themes and how frequently they appeared in interviews, with no reportable disagreements.  The analysis of interviews is presented in Section \ref{sec:Results}.

After completing informed consent, to address \textit{RQ2}, we conducted four 60-minute participatory design workshops, each with 1-2 clinicians and 1-2 researchers. Workshops began with an overview of the eight themes identified in prior research, which framed the working design space, followed by a brief on the design goal and constraints, which was to prototype games that secretly employ game controller input to guide autistic children in exercises that build motor skills deployable at home and in clinics.

Workshops continued with a warm-up activity \cite{marquez2021physical} to prepare bodies and minds for bodystorming \cite{marquez_segura_embodied_2016, marquez_segura_bodystorming_2016}. Then, clinicians were asked to describe playful activities they have already used in practice. Using these ideas and a Crazy Eights ideation activity, all members of the workshops quickly sketched and wrote eight ways these activities could be augmented with technology.

The remainder of the workshop followed a tight Situated Play Design structure \cite{altarriba_bertran_chasing_2019, altarriba_bertran_chasing_2020, altarriba_bertran_situated_2021} in which the earlier sketches were further explored through bodystorming, digital prototyping in Figma, and discussion of related existing technologies. Then participants deployed these prototypes by
physically acting out scenarios using embodied sketching methods \cite{marquez_segura_embodied_2016}. The workshop ended with a design debrief to refine the protocol \cite{zimmerman_recentering_2022, drachen_games_2018}. The low-fidelity and high-fidelity outcomes of the design workshops are presented in Section \ref{sec:Design}.

\section{Interview Themes (\textit{RQ1})} \label{sec:Results}
We identified eight prominent themes that describe clinicians' requirements for effective game-based interventions in pediatric physical and occupational therapy. 
 
\noindent\paragraph{Theme 1: Steps to Exercises}\label{theme1}Discussed in all interviews, the first identifiable theme appeared as the \textbf{importance of being able to break down exercises into smaller steps}. P02 explained:
\textit{
    "Running is hard, … but there's a lot of components that go into that… As a therapist, you'd say, ‘What do you need to run?’…For some kids, they really need to improve their balance … or their legs just aren't strong. Or they're not running with a pattern that helps. So it's kind of breaking that down." - P02}. Thus, effective interventions must support breaking complex motor goals into smaller, therapist-defined components.

\noindent\paragraph{Theme 2: Support Needs and Flexibility}\label{theme2}Therapists in all interviews emphasized that interventions must be continuously and quickly adapted to different individuals' support needs. As P03 explained:
\textit{
    "I definitely just think there are different levels of function when it comes to autism, and what is beneficial to some is not to others." – P03}.
Therapists reported that rigidly graded games lead to disengagement. P04 described abandoning an inflexible game.
\textit{
     "I stopped using it because I'd have kids try their hardest to make a letter, and unless it was perfect, they wouldn't get credit for it. So then a lot of kids didn't want to participate in it." - P04 }.

\noindent\paragraph{Theme 3: Games Lacking Therapeutic Specificity}\label{theme3}Six therapists (P02, P03, P04, P05, P07, P09 ) explained that games are often not tailored to the goals and needs of therapeutic practice. P07 noted how a game without therapeutic specificity could cause a child to focus on the wrong things:
\textit{
    "Kids just want to get a high score. Sometimes you can get the high score without doing the growth motor skills that we're looking for… like with [Wii] boxing, as long as they're moving their hands as fast as they can. They're getting a high score. But they're not getting the trunk rotation, core activation, and extension of the elbows that I'm looking for from a therapy perspective." - P07}.
Current games are not specifically tailored for pediatric therapy, but rather for entertainment, highlighting a discrepancy between existing games and specific motor outcomes targeted in therapy.

\noindent\paragraph{Theme 4: Accessibility and User Friendliness}\label{theme4} Prioritizing accessibility and user friendliness from the initial design phase was identified as essential by four therapists (P01, P04, P05, P09). P01 stated:
\textit{
"My only concern is that, since COVID, I've realized that internet access, even space within a house, is sometimes limited." - P01}.
Access to gaming consoles and technology is not universal. Thus, the modality of a game is essential to consider. 

\noindent\paragraph{Theme 5: Sensory Overload}\label{theme5} Therapists (P01, P04, P05, P06, P09) discussed the barrier imposed by the aesthetics of game-based interventions on their patients:
\textit{
    "It can be too overwhelming if there are too many things present…it might overstimulate them or stress them out, so it kind of just depends on the patient." - P09}. 
Therapists emphasized that \textbf{visual choices should be structured in a way that allows children to spend less effort processing the sensory environment of the game, and more time focusing on the content of the activity itself}. Additionally, the environment of an intervention should be adjustable to account for differences in sensory perceptions.

\noindent\paragraph{Theme 6: Lesson Carryover}\label{theme6} Therapists (P01, P04, P05, P07, P08) spoke on the difficulty of ensuring carryover outside clinical settings. Without practicing at home, skills introduced in therapy would fade.  P04 said:
\textit{
    "So we're trying to teach kids to tie their shoes, but if they don't have shoes that have ties [at home]. Then, a year later, they're like, ‘I don't remember.'" - P04}.

\noindent\paragraph{Theme 7: Choice and Autonomy}\label{theme7} All therapists emphasized that allowing children to make choices increased engagement in therapeutic activities. The act of choosing can incentivize engagement and motivation:
\textit{
    "Giving a field of 2 options, even if they don't necessarily like either. That's really effective, because they're still the ones choosing. It's giving them the choice. That's really important." - P04}.

\noindent\paragraph{Theme 8: Social Influence}\label{theme8} Many therapists (P01, P03, P07, P08, P09) spoke on how \textbf{doing an activity for a social goal shifts the activity from a clinical task} to a shared experience:
\textit{
    "It doesn't seem like homework if they get to do something else with others. Be it with peers or family, it gives the success of being able to do something others do, and other people wanting to do that activity with them." - P01}.
Sharing an activity with others reinforces engagement, builds confidence, and facilitates interactions with others. 

\section{Speculative Design Outcome of Participatory Workshops (\textit{RQ2})}\label{sec:Design}

\begin{figure*}[htbp]
    \centering
    \includegraphics[width=0.7\textwidth, trim=10 20 0 20,clip]{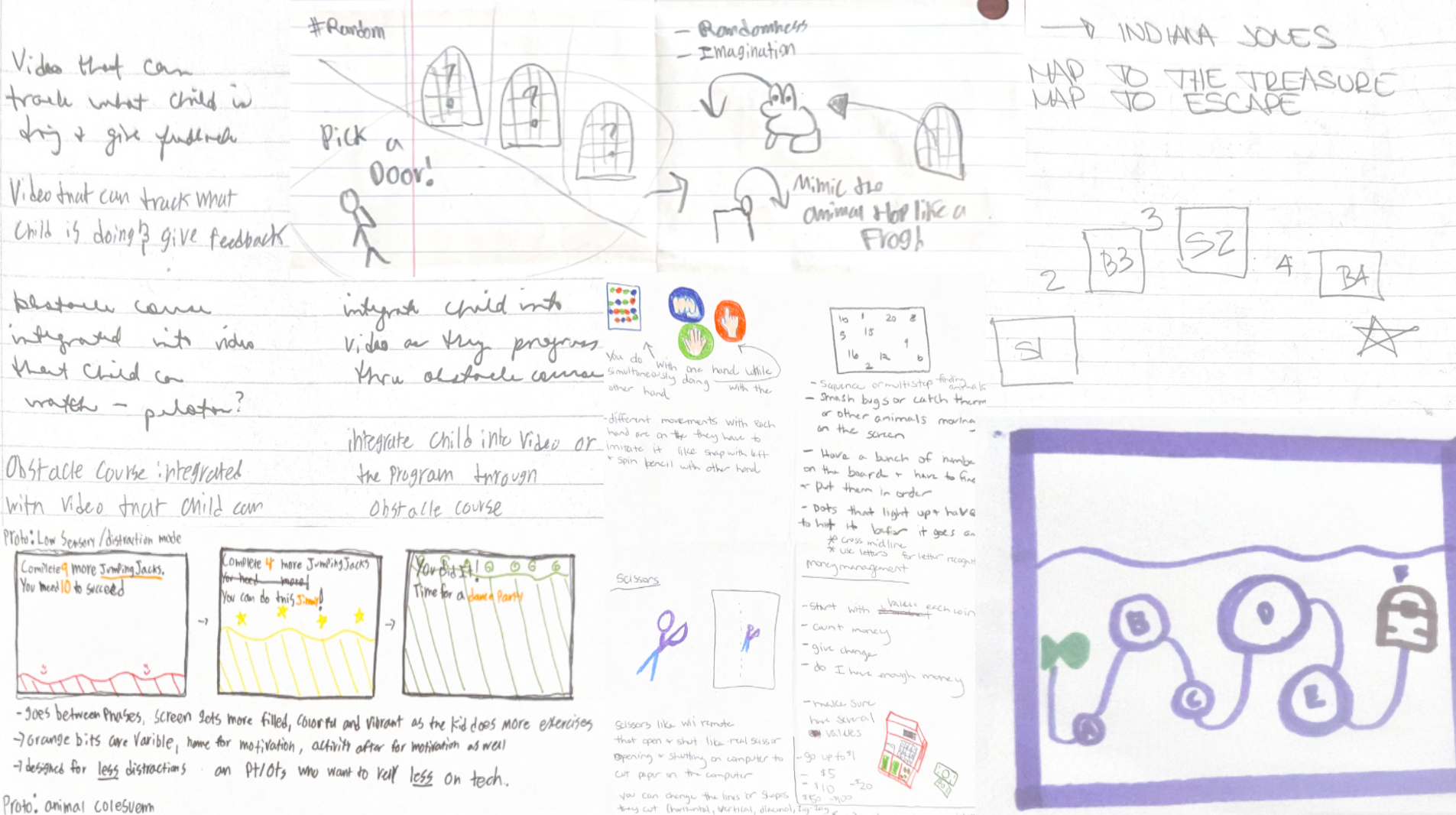}
    \caption{A collage of collected low-fidelity prototypes from our co-design workshops}
    \label{fig:LoFi}
    \Description{A collage of collected low-fidelity prototypes from our co-design workshops, some are text, some are sketches, some are idealized interfaces}
\end{figure*}

\begin{figure*}[htbp] 
\centering

\begin{subfigure}{\textwidth}
    \centering
    \includegraphics[width=0.85\textwidth, trim= 10 10 10 10,clip]{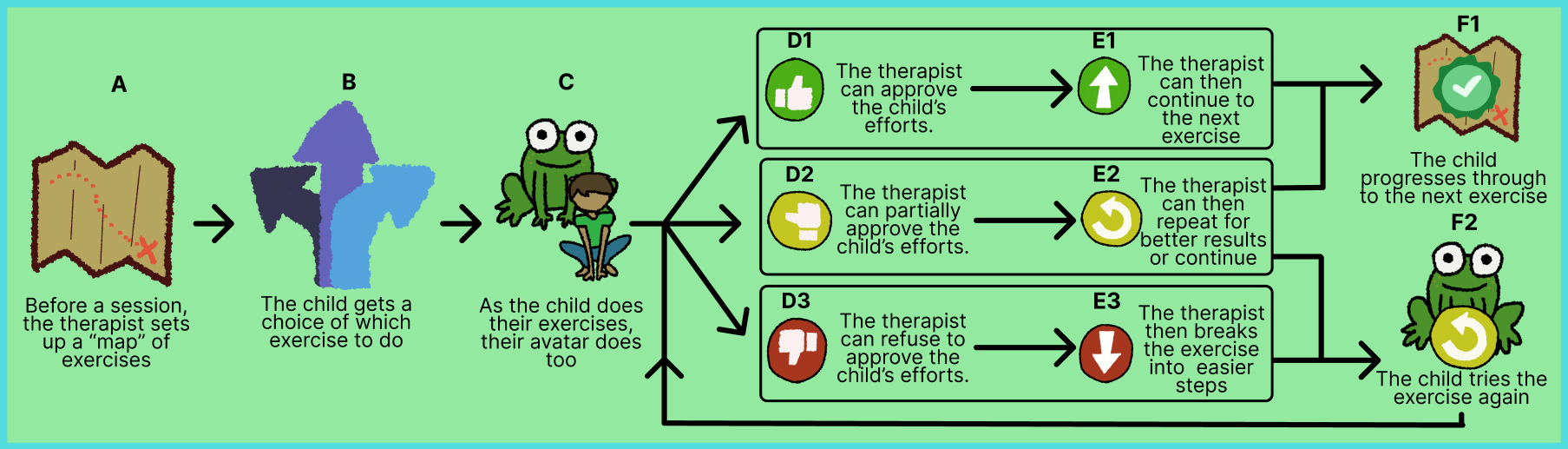}
    \caption{Flow Chart of Themes Implemented Into a Speculative Design}
    \label{fig:ConceptualPlatform}
    \Description{A conceptual flow showing how identified themes inform the design}
\end{subfigure}
\\
\begin{subfigure}{\textwidth}
    \centering
    \includegraphics[width=0.85\textwidth, trim=12 10 18 10,clip]{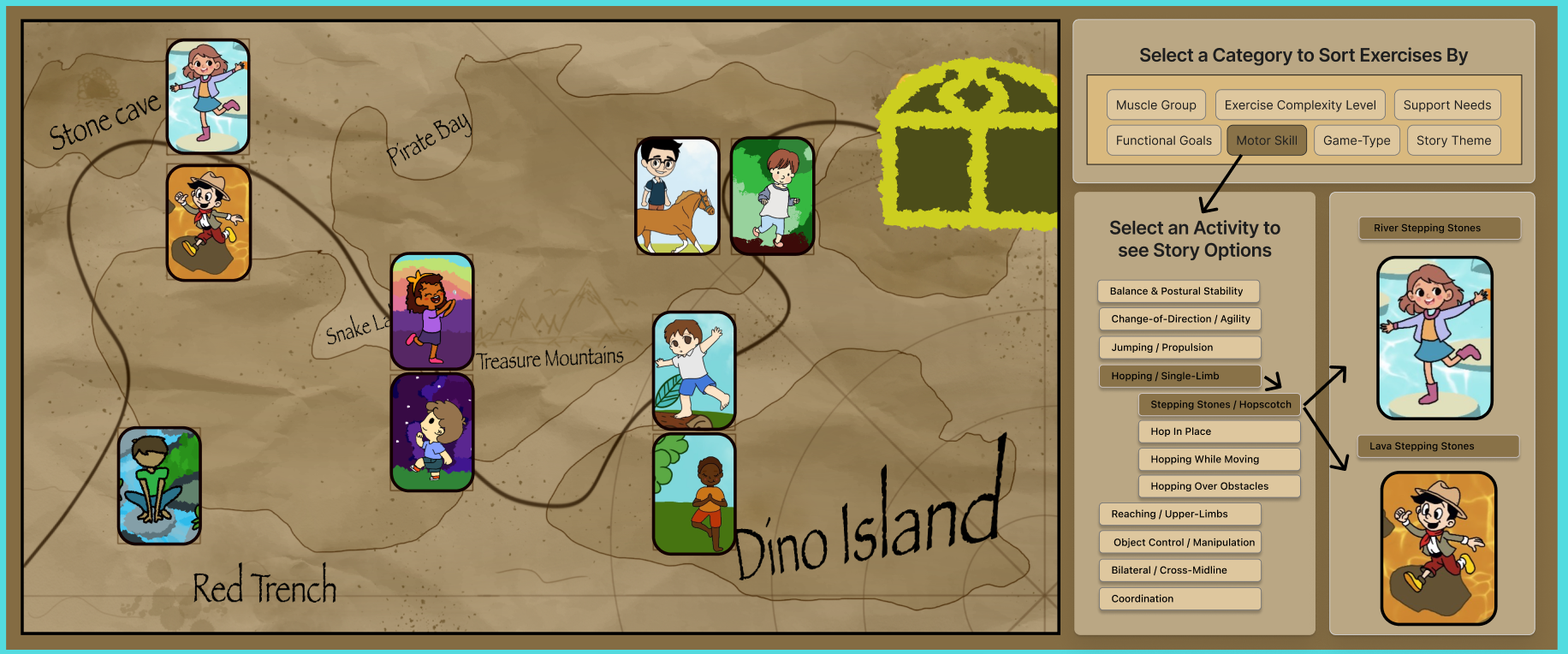}
    \caption{A Therapist's view of designing a "Map" of activities before a session}
    \label{fig:TherapistInterface}
    \Description{The therapist interface for creating customized activity sessions}
\end{subfigure}
\\
\begin{subfigure}{0.485\textwidth}
    \centering
    \includegraphics[height=0.16\textheight,keepaspectratio,trim=7 7 7 7,clip]{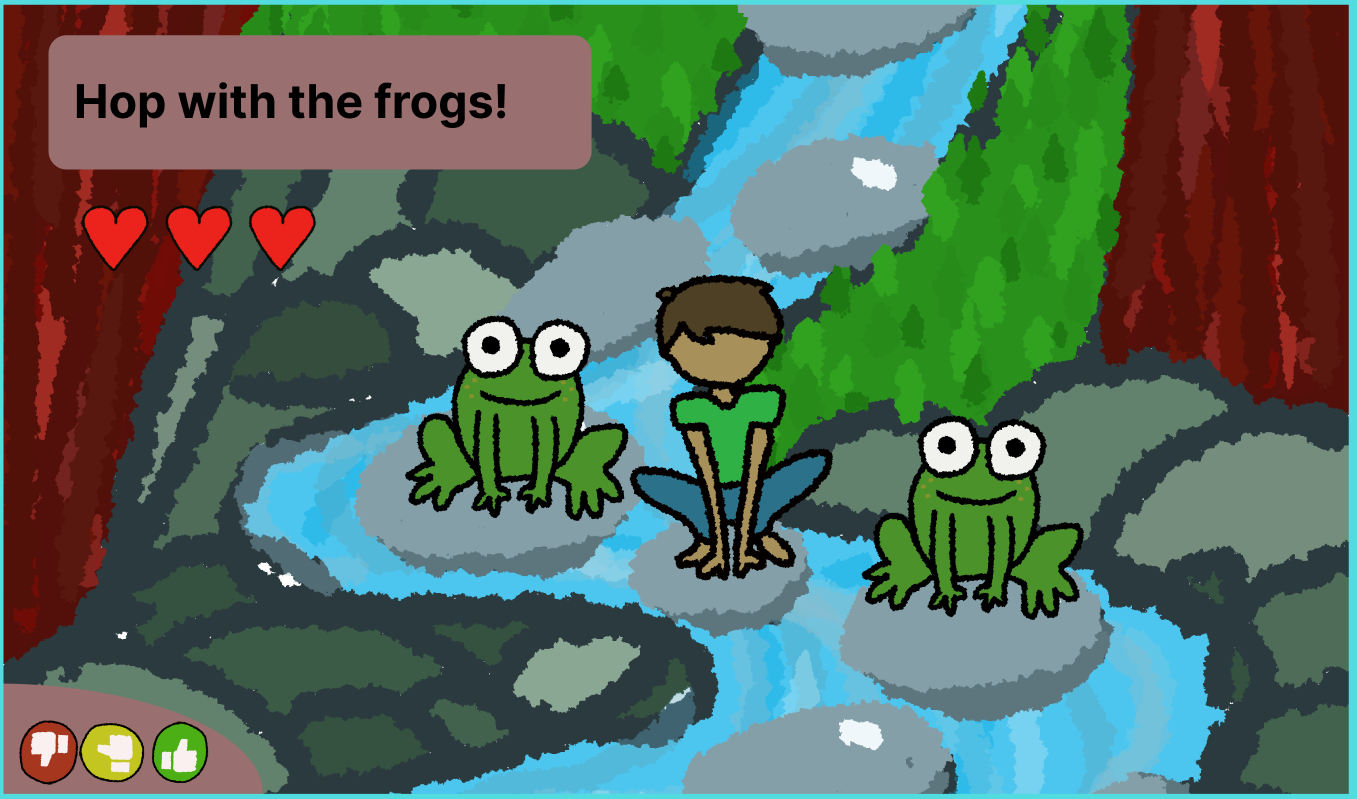}
    \caption{The screen the child sees as they do their exercises}
    \label{fig:PlayerInterface}
\end{subfigure}
~
\begin{subfigure}{0.485\textwidth}
    \centering
    \includegraphics[height=0.16\textheight,keepaspectratio,trim=7 7 7 10 ,clip]{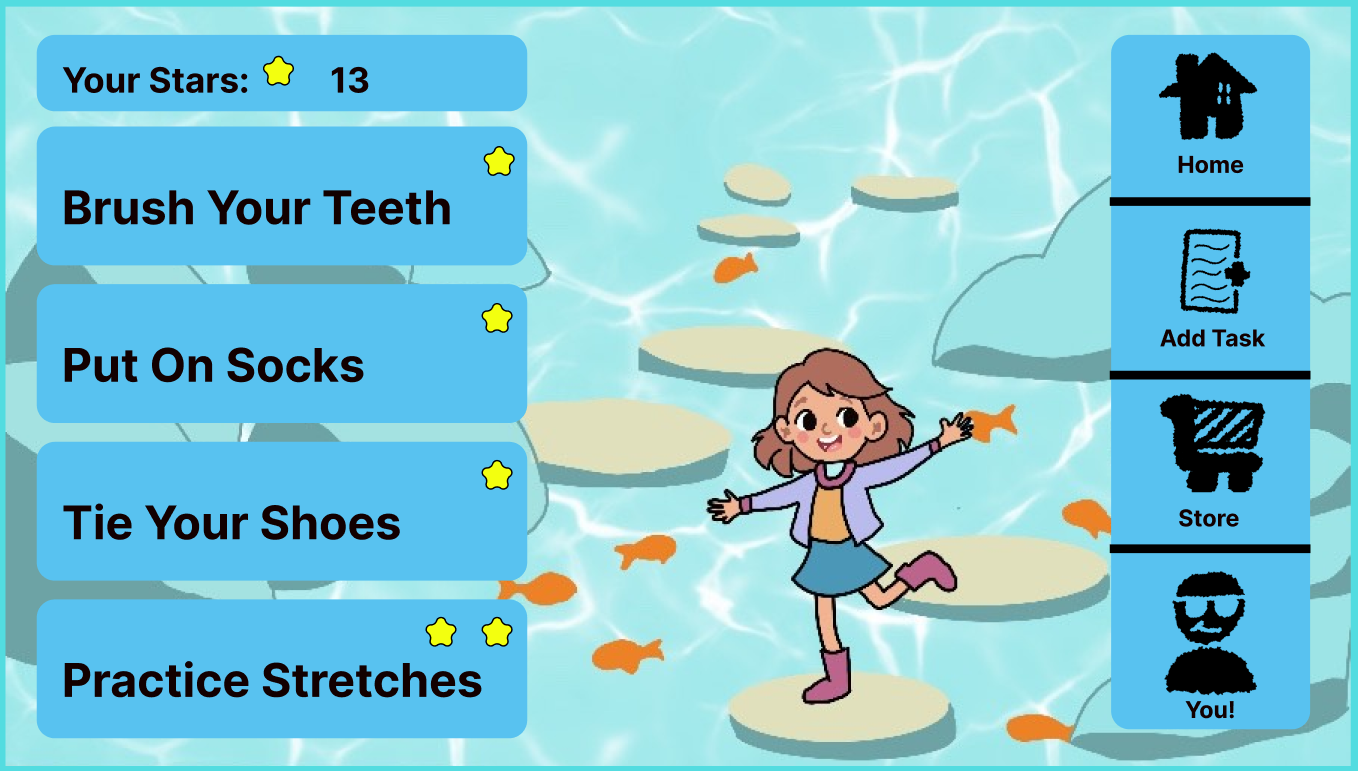}
    \caption{A Mock-up for at-home ADL/Motor goals}
    \label{fig:ParentInterface}
\end{subfigure}

\caption{Speculative design of a modular therapeutic gaming platform: (a) conceptual flow showing how identified themes inform the design, (b) therapist interface for creating customized activity sessions, (c) child-facing gameplay interface, and (d) caregiver interface for home use.}
\label{fig:SpeculativeDesign}

\end{figure*}

Inspired by our eight themes (Section \ref{sec:Results}) and our low-fidelity prototypes from participatory design sessions, which are available in Figure \ref{fig:LoFi} and our supplemental materials, we present the resulting speculative design in Figure \ref{fig:SpeculativeDesign} \cite{augerSpeculativeDesignCrafting2013}, based on the following design principles and adoption factors:

\begin{enumerate}[nosep]
    
    \item \textbf{Design Principles:} Our findings indicate that effective games for pediatric therapy must support stepwise task breakdown, therapist-driven flexibility, and therapeutic specificity, accessibility, sensory customization, lesson carryover, child autonomy, and social engagement. Most importantly, these requirements are not independent; they must be supported together in a single system to best assist clinical practice.
    
    \item \textbf{Adoption Factors:} All therapists expressed interest in games, but only if they allow for flexibility. That being said, identified barriers to adoption were a lack of flexibility, rigid assessments, technological complexity, and misalignment between in-game success and therapeutic goals.
    
\end{enumerate}

\noindent\paragraph{Steps to Exercises and Therapeutic Specificity}To allow exercises to be broken down into smaller steps, the speculative design allows the clinician to create a map of exercises before the session, allowing for therapeutic specificity. Prior work also encourages this, as there is importance to having incremental, play-based learning in clinical contexts \cite{Fiss2021}. By co-creating our speculative design with clinicians, we help avoid the barrier of many existing VGBIs that prioritize fun over therapeutic outcomes \cite{duval_designing_2022}. Figure \ref{fig:TherapistInterface} shows an interface for therapists to configure a map of exercises before a session.

\noindent\paragraph{Addressing Diverse Support Needs Through Flexibility}A central contribution of this design is how it works with therapist-driven evaluation rather than automation. Using a Wizard-of-Oz paradigm, therapists provide graded and therapeutically-specific feedback (e.g., approve, partial approve, retry) based on how the child does the exercise. This allows the therapist to dynamically consider factors and this ambiguity in grading helps our system accommodate a plurality of modes rather than enforcing a single standard grading system \cite{turmo_vidal_ambiguity_2024}. This flexibility also helps encourage different challenge levels, which therapists identified as critical for engagement, a concept aligning with flow theory \cite{csikszentmihalyi_flow_2014}. Figure \ref{fig:PlayerInterface} shows a gameplay interface for the child, where they can see their avatar exercise with them, and in the bottom left are the three grading buttons for the therapist.

\noindent\paragraph{Accessibility and Sensory Considerations} To address accessibility, user friendliness, and sensory customization, the speculative design was made to work with a phone-based interface that can be screen cast to a television, as this was the most popular technology among interviewees. Each exercise has an attached QR or link to a video of the exercise, to help show how to do it. The design includes a page of sensory customizations based on feedback from therapists, which can adjust visual and auditory information to prevent over-stimulation.

\noindent\paragraph{Autonomy and Social Engagement.} Given the important role of autonomy and choice, the in-game map is set up to allow the child to pick their next exercise. The maps are initially made by the therapist, but choices for children are integrated throughout the map, creating an engaging experience \cite{calvo_autonomy_2014}. To support social influence, the design anticipates allowing exercises to be completed cooperatively or competitively with others, through methods such as replaying exercises and taking turns to see who can get the best score, decided and graded by the therapist, or co-presence. Both options help frame the exercise as a fun shared experience, as opposed to an isolated clinical exercise.

\noindent\paragraph{Anticipating Lesson Carryover.}  Prior work suggests that motor learning is most effective when generalized across different contexts \cite{Levac2009}. The design anticipates having a secondary at-home mode, where a child can do safe exercises alone, and have their parent sign off on these exercises as done, or more difficult exercises with their caregiver acting as the therapist, and using the linked tutorial videos to run the exercise.  Figure \ref{fig:ParentInterface} shows a mock-up implementing carryover, where a child can complete simple tasks at home and unlock coins for different incentives.

\section{Discussion and Future Work}\label{sec:Discussion}
\subsection{Implications of \textit{RQ1}}
Our findings reveal that effective pediatric therapeutic games require interdependent design considerations: stepwise task breakdown, therapist-driven flexibility, therapeutic specificity, accessibility, sensory customization, lesson carryover, child autonomy, and social engagement must be supported holistically as foundational requirements. A key tension emerges between HCI's push toward intelligent automation and clinicians' need for interpretive control. Therapists explicitly rejected rigid automated assessment, describing how inflexible grading causes disengagement, inverting a common design assumption by suggesting designers ask ``how can we augment clinician judgment?'' rather than ``how can we automate feedback?'' This extends prior critiques of automation in care contexts \cite{elish_moral_2019, spielAgencyAutisticChildren2019, kayali_design_2018, baltaxe-admonyDREEMMovingEmpathy2024, duval_chasing_2021, morozovEverythingClickHere2013, charlton_nothing_2000, ymousAmJustTerrified2020, wongResistanceHopeEssays2018}. Our participants also revealed that the meaning of failure matters as much as its frequency: when therapists control feedback, imperfect motor output could be framed as progress rather than rejection and as a communicative act requiring careful authorship, building on flow theory \cite{csikszentmihalyi_flow_2014}.

\subsection{Implications of \textit{RQ2}}
Clinicians expressed cautious optimism but identified consistent adoption barriers: rigid automated assessment, technological complexity, and misalignment between in-game and therapeutic success, suggesting clinician control is the key adoption lever. Our speculative design uses Wizard-of-Oz not as a prototyping technique but as a permanent interaction paradigm \cite{dow_wizard_2005, duval2022playful}, where therapists provide real-time graded feedback rather than algorithms. This acknowledges that no sensor or ML model captures the contextual factors therapists weigh (fatigue, anxiety, genuine effort despite imperfect execution). Clinicians described constantly adapting interventions. Therapeutic games should be designed for appropriation rather than fidelity \cite{turmo_vidal_ambiguity_2024, jared_duval_reimagining_2023}, which is why we propose the three-button feedback system (approve, partial, retry) to deliberately embrace evaluative ambiguity.

\subsection{Limitations and Future Work}

While co-design with clinicians helped shape our concept, we emphasize that autistic children and their families must be central contributors in future design. Children are the primary users of the intervention and will be best suited for informing what is engaging, enjoyable, and meaningful in gameplay. Caregivers will help inform the design of the at-home version, as they will be the primary controllers for that mode. Our next phase of research will involve co-design sessions with autistic children and their families to help design the mini-game structures, themes, and rewards. Further work will also include iterative prototyping and evaluations of usability in clinical sessions.

\section{Conclusion}\label{sec:Conclusion}
This work examined pediatric physical and occupational therapists' experiences with game-based interventions to identify clinician-informed requirements for \textit{AutMotion Studio}. Through semi-structured interviews with nine clinicians, we identified eight interdependent themes that function as foundational requirements for clinical adoption, including stepwise task breakdown, flexibility for diverse support needs, therapeutic specificity, accessibility, sensory customization, lesson carryover, child autonomy, and social engagement. Clinicians emphasized that common barriers (rigid assessment, technological complexity, misalignment of game and therapeutic success) could be addressed through systems granting interpretive control, motivating our speculative design's Wizard-of-Oz paradigm. We contribute to HCI and accessibility research by foregrounding how ambiguity, flexibility, and human judgment can be intentionally designed into health interventions, offering an alternative to automation-first approaches in serious games for health.

\bibliographystyle{ACM-Reference-Format}
\bibliography{refs.bib}

\end{document}